\documentstyle[12pt]{article}
\begin{document}

\begin{flushright}
RU--98--33 \\
\today 
\end{flushright}

\begin{center}
\bigskip\bigskip
{\Large  The Discrete $Z_{2N_c}$ Symmetry And Effective Superpotential In   
SUSY Gluodynamics}
\vspace{0.3in}      

{Gregory Gabadadze}
\vspace{0.2in}

{\baselineskip=14pt
Department of Physics and Astronomy, \\
Rutgers -- The State  University of New Jersey \\
Piscataway, New Jersey 08855, USA} \\
{\rm \it email address: gabad@physics.rutgers.edu}

\vspace{0.2in}
\end{center}

\vspace{0.9cm}
\begin{center}
{\bf Abstract}
\end{center} 
\vspace{0.3in}
We find  an expression for the effective superpotential 
describing the $N_c$ vacua of $SU(N_c)$ 
SUSY gluodynamics.  The superpotential       
reduces in some approximation to 
the Veneziano-Yankielowicz  expression  amended by the 
term restoring the  discrete $Z_{2N_c}$ symmetry. 
Moreover, the superpotential, being restricted  to 
one particular vacuum state, 
yields the expression which was derived recently to describe 
all the lowest-spin physical states  of the theory. 
The corresponding scalar potential has no cusp singularities 
and  can be used to study the domain walls interpolating
between the chirally asymmetric vacua of the model.  
 
\newpage
\noindent 
{\bf  Introduction} 
\vspace{0.2in}

Supersymmetric gluodynamics, the theory of gluons and gluinos,
seems to be an extremely useful testing ground for various
nonperturbative phenomena occuring in conventional QCD.  
The Witten index of the $SU(N_c)$ SUSY gluodynamics
equals to $N_c$ \cite {WittenN}. Thus, the ground
state of the model consists of at least $N_c$ different 
vacua parametrized by the imaginary phase of a nonzero gluino condensate 
\cite {WittenN}, \cite{Condensate}. The different 
vacua are related by discrete $Z_{2N_c}$ transformations
of gluino fields. Once one of the $N_c$ vacua is 
chosen, the $Z_{2N_c}$ symmetry group spontaneously breaks down   
to the $Z_2$ subgroup. As a result of the  discrete symmetry
breaking one expects to find domain walls separating the $N_c$
vacua of the model. 

Recently, Dvali and Shifman found that
the $N=1$ SUSY algebra admits some central extension if 
domain walls are present in the model \cite {DvaliShifman1}, \cite
{DvaliShifman2}. Thus, the domain walls saturating the BPS bound
for the wall surface energy density might exist in the theory  
\cite {DvaliShifman1}, \cite {DvaliShifman2}. 

There are  yet another attractive
arguments why BPS domain walls should be present in the model. Recently, 
$N=1$ SUSY gluodynamics was 
realized \cite {WittenBranes} as a low-energy 
field theory emerging in a particular 
brane setup within the $M$ theory framework. 
In that picture  the domain walls 
can be regarded as  higher dimensional D-branes wrapped around 
some compactified dimensions \cite {WittenBranes}. The D-branes, being
extended objects  
on which open strings can end in string theories \cite {Dbranes},
can also be viewed as BPS solitons in  corresponding
low-energy theories of supergravity \cite {Solitons}.
Thus, in accordance with the Witten's construction \cite
{WittenBranes} the $N_c$ vacua of SUSY gluodynamics should be  separated
by the BPS domain walls on each of which color flux tubes (strings) can end 
\cite {WittenBranes}. This picture of the vacuum is quite 
attractive from the theoretical perspective as well as  from the point of 
view of lattice simulations of SUSY Yang-Mills  model where some indirect
signatures of this construction could  be observed \cite {Lattice}.

The most straightforward way to study the vacuum structure 
would be to find explicitly the domain wall solutions (for recent
reviews see Refs. \cite {ChibisovShifman}, \cite
{ShifmanTalk}). For that purpose  one needs to have an
effective action describing the $N_c$ vacua of the model.

The effective action for $N=1$ SUSY Yang-Mills  (SYM) was
proposed by
Veneziano and Yankielowicz (VY) \cite {VY}. 
The VY superpotential reproduces explicitly
all the quantum anomalies of SUSY gluodynamics.
However, it does not respect the 
discrete $Z_{2N_c}$ symmetry \cite {ShifmanKovner} which is 
left once the chiral $U(1)_{\rm R}$ invariance  
is broken by the axial anomaly. 

In order to restore $Z_{2N_c}$ invariance
the VY superpotential  was amended by an additional term \cite {ShifmanKovner}.
The resulting expression is $Z_{2 N_c}$ symmetric. However,
the corresponding scalar potential possesses cusp singularities
\cite {ShifmanKovner}. These cusps  are encountered
in the field space 
as one interpolates between the $N_c$ vacua \cite {ShifmanKovner}.
For that reason the amended VY superpotential  can not be used 
to describe the domain walls separating chirally asymmetric vacua \cite {KKS}.
Moreover, considering SUSY YM with some heavy matter multiplets added 
(i.e. SUSY QCD with heavy flavors)
and  gradually integrating out those heavy states, 
one shows that the domain walls of the chirally asymmetric vacua 
cannot be found within the VY framework \cite {Smilga}\footnote[4]{
This assertion is valid  modulo some assumptions on the 
vacuum structure and form of the K\"ahler 
potential, see discussions in Ref. \cite {Smilga1}.}. 

On the other hand, recent studies \cite {GiaZura} of the model which shares
in the large $N_c$ limit some 
important features of SYM  manifestly demonstrated the existence of BPS  
domain walls with the  properties required in the brane construction
\cite {GiaZura}. 

Putting the whole  set of arguments together one naturally concludes
that it must be the VY framework which does not account adequately for 
all properties of the complicated ground state.

There is yet another reason  to believe that the VY superpotential
is not complete. In Refs. \cite {FGS1}, \cite {FGS2} it was shown that
in order to account for all the lowest-spin excitations of the
model, one necessarily needs to introduce
an additional chiral superfield in the VY description \cite {FGS2}.  

The aim of this work  is to use this additional chiral superfield
to  find an expression for the superpotential
which would respect the $Z_{2N_c}$ invariance.  The superpotential
should lead as well to 
the scalar potential with no cusps. Moreover, 
in some approximation the superpotential should 
reduce to the known expression  of
Ref. \cite {ShifmanKovner}. Finally, once restricted 
to some particular vacuum state it should yield the superpotential
derived recently in Refs.  \cite {FGS1}, \cite {FGS2}. 

In the next section we show that such a 
superpotential can really be found. 
\vspace{0.3in}\\
1. {\bf  The Discrete Symmetry and VY Superpotential}
\vspace{0.2in}

The classical action of  $N=1$ SYM theory is invariant  
under chiral, scale 
and superconformal transformations. 
Once quantum effects are taken into account, 
these symmetries are 
broken by the  chiral, scale and superconformal  anomalies respectively.
Composite operators that appear  in the expressions for the 
anomalies can be gathered into   
a composite chiral supermultiplet ${\rm Tr} W^{\alpha}W_{\alpha}$ 
\cite {WessZumino} (we use the notations of \cite {WessBagger})

The effective action of the model can be a functional of the
superfield $S$ 
\begin{eqnarray}
S\equiv { \beta (g) \over 2 g} \Big \langle 
{\rm Tr} W^{\alpha}W_{\alpha}\Big \rangle _Q\equiv
A(y)+\sqrt{2} \theta \Psi(y) + \theta^2 F(y),  \nonumber
\end{eqnarray}
where the VEV is defined for nonzero value of an external 
(super)source $Q$ \cite {Shore}.
$\beta(g)$ stands for   the SYM beta function which is known 
exactly \cite {beta}. The lowest component
of the $S$ superfield $A$ is bilinear in gluino fields and has the quantum
numbers of the scalar and pseudoscalar gluino-gluino  bound states.  The
fermionic component in $S$ is related to  the gluino-gluon composite and
the $F$ component of the chiral superfield includes operators
corresponding to both the scalar and 
pseudoscalar glueballs ($G_{\mu\nu}^2$ and 
$G_{\mu\nu}{\tilde G^{\mu\nu}}$ respectively) \cite {VY}. 

Assuming that the effective action (more precisely,  
the generating functional for one-particle-irreducible (1PI) 
Green's functions \cite {GoldstoneSalamWeinberg}) of the 
model can be written in terms of the single  superfield $S$,   and 
requiring also that the effective action respects all the global
continuous symmetries and 
reproduces the anomalies
of the SYM theory,   one derives the Veneziano-Yankielowicz 
effective superpotential \cite {VY} 
\begin{eqnarray}
{\cal W}_{\rm VY}(S)=
\gamma ~S~ {\rm ln} {S\over e \mu^3},   
\label{VYsup}
\end{eqnarray}
where $\gamma \equiv - (N_c g/16 \pi^2 \beta(g))>0$, $\mu$ stands for
the dimensionally transmuted scale of the model and $e\simeq 2.71$. 

It was noticed in ref. \cite
{ShifmanKovner} that the VY action does not respect the discrete
$Z_{N_c}$ symmetry -- the nonanomalous remnant of anomalous 
$U(1)_R$ transformations\footnote[3]{The actual discrete
symmetry group of fermion field  transformations  is $Z_{2N_c}$. Since  
all the quantities below will be written in terms of fermion bilinears
the symmetry reduces to  $Z_{N_c}$.}.
In order to make the action invariant under $Z_{N_c}$
transformations the VY superpotential was amended in Ref. 
\cite {ShifmanKovner} by the following term 
\begin{eqnarray}
\Delta {\cal W}=
i \gamma ~ {2\pi n \over N_c}~S ,   
\label{KSsup}
\end{eqnarray}
where $n$ enters in the action as an integer-valued Lagrange multiplier.
The partition function of the theory should be regarded as a 
sum of path integrals where $n$ runs from $-\infty$ to
$+\infty$ \cite {ShifmanKovner}.

Thus, after the term (\ref{KSsup}) is included the action 
becomes $Z_{N_c}$ invariant \cite {ShifmanKovner}.  
The ground state of the model consists of at least $N_c$ different vacua
labeled by different values of the phase of the gluino condensate. 
The resulting scalar  potential which  respects the discrete $Z_{N_c}$
symmetry can be written as \cite {ShifmanKovner}, \cite {KKS}
\begin{eqnarray}
U(\phi) \propto (A^* A)^{2/3} {\rm ln}({A^*\over \mu^3}e^{i2\pi n/N_c}) 
{\rm ln}({A\over \mu^3} e^{- i 2\pi n/N_c}), 
\label{glued}
\end{eqnarray}
where 
\begin{eqnarray}
{(2n -1)\pi\over N_c} < {\rm arg}(A) < {(2n+1)\pi \over N_c}.
\nonumber
\end{eqnarray}
Thus, the complex plane of ${\rm arg}(A)$ is divided 
into $N_c$ sectors. The potential is continuous
in the  plane, however it has cusps 
at ${\rm arg}(A)=(2n +1)\pi/N_c$ \cite {ShifmanKovner}. 

If one is restricting the superpotential to one particular vacuum state
with some definite value of the phase of the gluino condensate,
then the expression should account for all possible low-energy
degrees of freedom of the theory. 
In Refs. \cite {FGS1}, \cite {FGS2} it was argued that the 
VY Lagrangian should be modified further in order to include 
all the lowest-spin low-energy degrees of freedom of the $N=1$ SUSY YM
model. In fact, it was
shown that to account for  glueballs 
the effective superpotential should be defined in terms of two 
chiral supermultiplets \cite {FGS2}. The supermultiplet $S$ 
in that  construction includes fields with quantum numbers 
of gluino-gluino ``mesons'' (along with the fermionic gluino-gluon state)
while another chiral supermultiplet is needed to incorporate glueball states
\cite {FGS2}
\begin{eqnarray}
{\cal W}={\cal W(S)}_{\rm VY}+{\cal W}_1(S,~
{\rm Another~~Chiral~~Superfield}). 
\label{werbal}
\end{eqnarray}
Thus, the  second chiral superfield is 
needed to describe glueballs  as excitations over one particular
vacuum state \cite {FGS2}. In this respect,  
it  would be  nice to have that same 
superfield also restoring the  discrete
$Z_{N_c}$ symmetry which is lost in the VY superpotential. 

If this possibility is really realized, then the integer-valued Lagrange
multiplier term (\ref {KSsup}) should be occuring 
once the new chiral superfield  in (\ref {werbal}) is integrated out. 
Some examples of this type were discussed in Ref. \cite {KKS}.

Hence, our goal  
is to find out an  expression for the  
superpotential ${\cal W}$ as a function of two chiral superfields $S$
and  let us
say $X$ which would satisfy to the following requirements:
\begin{itemize}

\item The superpotential should be a homomorphic function of
arguments; 

\item  The superpotential
should reproduce all the anomalies of the model, i.e. it should 
contain the VY superpotential as an ingredient \cite {VY};

\item It should be invariant under the discrete $Z_{N_c}$
transformations \cite {ShifmanKovner};

\item The scalar potential should  have at least $N_c$
minima with broken chiral invariance;

\item If the superfield $X$ is integrated out, the superpotential 
should  yield the expression (\ref {VYsup}) amended by the term 
(\ref {KSsup});

\item If the superpotential is restricted to one particular
minimum with broken chiral symmetry, it should reproduce 
the generalized Veneziano-Yankielowicz 
superpotential derived in Refs. \cite {FGS1}, \cite {FGS2}. 
\end{itemize}

We would like to argue that such a superpotential exists.
The general form of the superpotential will be given in the next
section. Here, we consider a simple expression.
It can be  obtained as a part of the general solution and 
should be regarded as a  toy example used to elucidate   
the construction. 

One defines 
\begin{eqnarray}
{\cal W}(S, X)\equiv \gamma~ S~ {\rm ln}{S\over e \mu^3}    
+\gamma ~S~\Big ( X- {1\over N_c}~{\rm sinh}(N_c X) \Big ),
\label{super}
\end{eqnarray}
where the first term is nothing but the VY superpotential
(\ref {VYsup}). The second term is supposed to restore 
the discrete $Z_{N_c}$ invariance of the VY superpotential. Notice that
$X$ is a dimensionless chiral superfield with zero $R$ charge.  

Let us now check whether the expression (\ref {super}) really 
satisfies to all the requirements listed above. 
First of all, the expression (\ref {super}) 
yields all the anomalies 
of the model;  indeed, the first term in 
 (\ref {super}) is just the VY superpotential which is 
designed to reproduce correctly the anomalies \cite{VY}.
The second term does not contribute to the anomalies.  
 
Consider the discrete $Z_{N_c}$
transformations. The chiral superfield $S$ transforms as 
$$
S \rightarrow {\rm exp}\Big (i {2\pi k\over N_c} \Big)~S, 
~~~~~k=0, 1,..., N_c-1.
$$ 
As a result of this transformation the first term in the 
expression (\ref {super}) and its conjugate generate  an additional 
term in the Lagrangian. This term has the form 
$$
i{2\pi k\over N_c}~\gamma~ (S|_F - S^+|_{F^+}). 
$$
This expression can be eliminated by the following 
shift  of the $X$ superfield
\begin{eqnarray}
X\rightarrow X- i{2\pi k\over N_c}. 
\label{Shifts}
\end{eqnarray}
However, there is a restricted class of possible shifts which one 
is allowed  to perform  in the partition function of the model. 
The shifted fields,  which along with the initial fields 
are being considered as physical ones, 
should also satisfy appropriate boundary conditions. In other words,
the shifts we are discussing 
should be transforming the $X$ field 
from one vacuum state to another. Anticipating the results of our
discussions below, the vacuum values of the $X$ field are just 
going to be multiples of $i2\pi/N_c$, thus the shifts  (\ref{Shifts})
do satisfy to the requirements set above. 
Hence, the superpotential (\ref {super}) really respects the discrete
$Z_{N_c}$ invariance\footnote[2]{One should also make sure that
the K\"ahler potential of the model is invariant w.r.t. the
shifts  (\ref{Shifts}). See the discussion of the K\"ahler potential
in Section 3.}. 

Let us now check what happens if one naively integrates out the $X$
field from the expression (\ref {super}) (though, there is no physical
reason to do that). The equation for 
$X$ minimizing the scalar potential
is given as\footnote[3]{In deriving the scalar potential one should 
actually switch to chiral superfields with an appropriate dimensionality 
$\Phi\equiv S^{1/3}$ and $Y\equiv X \Phi $ 
and calculate minima w.r.t. those  superfields (in this particular
case the answer is the same, see below).}
$$
{\partial {\cal W}(S, X)\over \partial X}=\gamma~S~ \Big (1-{\rm cosh}(N_cX)
\Big )=0.
$$
Solving the equation  for $X$  (at nonzero  $S$ ) one finds
$$
X_*=i{2\pi n\over N_c}, ~~~~~~~~n=0,\pm 1, \pm 2,....\pm \infty.
$$
Substituting this identity  back into the superpotential (\ref
{super}) we derive 
\begin{eqnarray}
{\cal W}(S, X_*)=\gamma S~ {\rm ln}{S\over e \mu^3}+
i \gamma  {2\pi n\over N_c}~ S.
\label{KS}
\end{eqnarray}
This expression is nothing but the VY superpotential (\ref {VYsup})
amended by the term (\ref {KSsup}) of Ref. \cite {ShifmanKovner}. 
Thus, the term (\ref {KSsup}) is obtained if the $X$ field is
being integrated out. Later we will argue that the components of $X$
are related to glueballs. This excitations turn out to be 
lighter than the excitations described by the $S$ superfield \cite 
{FGS1}, \cite {FGS2}, so there is no physical reason to regard the
components  of the $X$ field as being integrated out. Thus, one should
keep the $X$ field in the superpotential as a necessary ingredient. 
 
The next step is to check whether 
the expression (\ref {super}) produces the  scalar potential with an 
appropriate  $Z_{N_c}$
structure. Let us introduce the following notations
$$
\Phi \equiv S^{1/3},~~~~~Y\equiv X~\Phi.
$$
Also, let us denote the  components of the superfields $\Phi$, $X$ and  
$Y$ as 
$\phi$, $\phi_x$ and $\phi_y$ respectively\footnote[4]{The change of
variables from $S$ to $\Phi$ is nonsingular in our case since we are 
dealing only with the phase of the theory where the VEV of the lowest
component of $S$ is nonzero.}.
The superfields
$\Phi$ and $Y$ have  right physical dimensionality.
The scalar potential $V$  can be written as a sum  of two terms 
$$
V(\phi,\phi_y)=V_1(\phi,\phi_y)+V_2(\phi,\phi_y), 
$$
where 
\begin{eqnarray}
V_1(\phi, \phi_y)\propto \Big | {\partial {\cal W}(\phi, \phi_y)\over 
\partial \phi} \Big |^2= \\ \nonumber
9\gamma^2~ |\phi|^4~ \Big |
{\rm ln}{\phi^3 \over \mu^3}+\phi_x - {1\over N_c}{\rm sinh}(N_c\phi_x)
-{\phi_x\over 3}[1- {\rm cosh}(N_c\phi_x)] \Big |^2,
\label{one}
\end{eqnarray}
\begin{eqnarray}
V_2(\phi, \phi_y)\propto \Big | {\partial {\cal W}(\phi, \phi_y)\over 
\partial \phi_y} \Big |^2=\gamma^2~ |\phi|^4~ \Big |1-{\rm
cosh}(N_c\phi_x)
\Big |^2. 
\label{two}
\end{eqnarray}
In these equations the substitution $\phi_y=\phi_x \phi$ is used
\footnote[5]{In the expressions (8) and (9)
the exact proportionality coefficients are set by the inverse
metric defined as the second derivative of the K\"ahler 
potential (see Ref. \cite {WessBagger}).}.

After the potential is set one can  list all the vacuum states of the
model. All those configurations should  satisfy
to the equations $V_1=V_2=0$. As one expects,  there 
are $N_c$ different vacua with broken chiral symmetry:
$$
|\phi|=\mu, ~~{\rm Re}\phi_x=0,~~{\rm Im}\phi_x=-3{\rm arg}\phi =
{2\pi k\over N_c},~~~k=0, 1,...N_c-1.
$$
These vacua differ from each other by the value of the phase
of the gluino condensate
\begin{equation}
\langle \lambda \lambda \rangle_k \propto  \mu^3 ~
{\rm exp}\Big (i{2\pi k \over N_c}\Big)~,~~~~~~~ k = 0,1,...,N_c-1~.
\label{condensate}
\end{equation}
Interpolating from one vacuum state to an another one 
no cusp singularities are encountered in (8) and (9).
The presence of the chiral
field $X$ smoothes out cusp singularities emerging in the case 
when $X$ is being integrated out\footnote[6]{
One also finds that there is a vacuum state with the zero value of
$|\phi|$. The existence of the
vacuum state with no gluino condensate was conjectured in Ref.
\cite {ShifmanKovner}. The question whether this phase of the model
can  actually be realized in the fundamental theory is a subject of recent
discussions \cite {SZ}, \cite{SM}, \cite {KKS}. In this work 
we concentrate on the vacua with the nonzero gluino condensate only.}.  

The next question we would like to elucidate is the physical
interpretation of the new chiral superfield. 
The $S$ superfield is related to the operator 
${\rm Tr} W^2$. The lowest component of
$S$ can be thought of as an interpolating field for a gluino-gluino
bound state. The question is whether one can find 
analogous identifications for the components of the new chiral
superfield. In order to clarify this question let us recall some
results of Ref. \cite {FGS2}. 
In Ref. \cite {FGS2}  an effective action 
describing physical excitations of the 
SUSY YM model in one of  the $N_c$  
vacua was constructed. The superpotential of Ref. \cite {FGS2}
is also written in terms of two chiral superfields, 
$S$ and some chiral superfield $\chi$. The first superfield 
was shown to describe gluino-gluino and gluon-gluino 
bound states, while the second one
was needed to include pure gluonic, glueball  states into the description. 
We shall argue below that the superfield $X$ 
in (\ref {super}) is also related to 
glueballs and is just a necessary ingredient of the 
effective superpotential (\ref {super}).  
In other words, 
we are going to show here that the superpotential (\ref {super})
reproduces the expression of Ref. \cite {FGS2} in the limit when 
one is restricted to some  particular chirally asymmetric vacuum. 
To accomplish this task  let us introduce the following notation:
\begin{equation} 
\chi \equiv 16 \gamma  \Big ( X- {1\over N_c}{\rm sinh} (N_cX) \Big ).
\label{chi}
\end{equation}
The $X$ field is a dimensionless chiral superfield and so is $\chi$. 
One  rewrites  the second  term in the 
superpotential (\ref {super}) in the following form:
\begin{equation} 
{1\over 16}~\chi~ S={1\over 16 N_c} \sum_{k=0}^{N_c-1}~\chi~ 
\Big (~S-\langle
S\rangle _k~\Big ),
\label{sum}
\end{equation}
where 
$$
\langle S\rangle _k \equiv \mu^3 {\rm exp}\Big (i {2\pi k\over N_c}\Big).
$$
Using these identities the superpotential
(\ref {super}) can be presented in yet another very useful form 
\begin{equation} 
{\cal W}(S, \chi(X) ) = \gamma~ S~ {\rm ln}{S\over e \mu^3}   
~+~{ \chi (X)  \over 16 N_c}~\sum_{k=0}^{N_c-1}~\Big (~S-\langle
S\rangle _k~\Big ).
\label{supersum}
\end{equation} 
The expression (\ref {supersum}) makes it transparent  
that the superpotential we are discussing can be obtained as a sum
of superpotentials defined for each particular  
$N_c$ vacuum state. Each of this vacua are labeled by the VEV of the 
gluino condensate with an appropriate phase. The initial  $Z_{2N_c}$
symmetry is spontaneously broken  down to $Z_2$ 
in each of these vacuum states.  
It is straightforward to determine how
the expression (\ref {supersum}) looks like when   
one restricts  consideration to some  particular vacuum state only. 
In that case one assumes that
the VEV of the $S$ field takes a single value, let us say for
simplicity $\langle
S\rangle _k
=\mu^3$. Then the expression (\ref {sum}) reduces to the 
following formula
$$
{\chi(X) \over N_c}~ \sum_{k=0}^{N_c-1}\Big  (S-\langle
S\rangle _k~\Big )\rightarrow \chi(X)~\Big  (~S-\mu^3~ \Big ).
$$
Substituting this expression back into Eq. (\ref {supersum})
one derives the following superpotential
\begin{eqnarray}
{\cal W}(S, \chi(X))\equiv \gamma~ S~{\rm ln}{S\over e \mu^3}    
~+~{1\over 16}~\chi(X)~\Big (~S-\mu^3~\Big ).
\label{FGS}
\end{eqnarray}
This is exactly the superpotential obtained  in Ref. \cite {FGS2}.
It describes the vacuum state of the model with broken chiral
invariance where the phase of the gluino condensate equals to
zero.  The initial  
$Z_{ N_c}$ symmetry in this vacuum is broken down to $Z_2$.  

Let us now return to our original question about the physical
interpretation of the components of the $X$ field. 
The components of the  $X$ field are related to 
the components of $\chi$ (Eq. (\ref {chi})).
On the other hand, 
the $\chi$ field is related to
glueball excitations of the $N=1$ SUSY YM model \cite {FGS2}. 
Thus, the components
of the $X$ field should also be related to the VEV's of the 
$G_{\mu\nu}^2$ and $G_{\mu\nu}{\tilde G_{\mu\nu}}$ composite operators
\footnote[6]{The explicit derivation is cumbersome and can be done using some
relations between $\chi$ and the tensor supermultiplet  used  in Refs.
\cite {FGS1}, \cite {FGS2}. For  the completeness of discussions 
we present here simple approximation results of  Refs. 
\cite {FGS1}, \cite {FGS2}. For 
the lowest component $\phi_{\chi}$ 
~~$
\partial^2 {\rm Re}\phi_{\chi} \simeq -{1\over 8 \mu^2}{\beta(g)\over
2g} \langle G_{\mu\nu}^2\rangle_{\rm s}$  $
\partial^2 {\rm Im }\phi_{\chi} \simeq {1\over 8 \mu^2}{\beta(g)\over
2g} \langle G_{\mu\nu}{\tilde G_{\mu\nu}} \rangle_{\rm p}, 
$
where the VEV's are functions of  some external fields  ${\rm s}$
and ${\rm p}$.
Thus, the real part of $\phi_{\chi}$ bears quantum numbers
of a scalar $0^{++}$ pure gluonic state, while the imaginary part is 
associated with 
the pseudoscalar $0^{-+}$ glueball state \cite {FGS1}, \cite {FGS2}.}. 

Thus, if one is restricted to study physics about 
some particular vacuum state
of the model, then the information about the whole $Z_{N_c}$ structure is
lost. As a result, the $\chi$ field can be introduced in accordance
with Eq. (\ref {chi}), and all the physical excitations about that
ground state can be described  in terms of  components of
the $\Phi$ and $\chi$ multiplets\footnote[9]{The statement that
the components of the $\chi$ multiplet are related
to glueballs assumes that the corresponding
K\"ahler potential is a function of the sum  $\chi+\chi^+$, see the
detailed discussions below.}. 

In our discussions 
we could have started from the generalized VY superpotential
of Ref. \cite {FGS2} and derived the $Z_{N_c}$ symmetric 
superpotential (\ref {supersum}) (and (\ref {super})).
Indeed, the superpotential (\ref {FGS})
describes physics of only
one particular ground state with a definite value of the phase of the
gluino condensate \cite {FGS2}. 
If one would need to generalize that expression to
include all the possible vacua, 
one should have summed the expression (\ref {FGS}) w.r.t. the 
phase labeling all the vacua. 
The result of this summation is given in  Eq. 
(\ref {sum}), the whole sum reduces to the quantity $\chi~S/16$. 
Thus,  one would have arrived at  the superpotential
(\ref {super}) written in terms of the $\Phi$ and $\chi$ fields
(without any knowledge about the $X$ superfield). 
Obviously, the question is how would one discover  in this approach 
that there is a substructure relating the $\chi$ field to the $X$ 
field (as in  Eq. (\ref {chi}))? The 
structure (\ref {chi}) would emerge as a result 
of the $Z_{N_c}$ symmetry 
and also supersymmetry itself. Indeed, if 
the superpotential (\ref {supersum}) is written in terms of the $\Phi$
and $\chi$ fields alone, and if the $\chi$ field is regarded as some
fundamental field, then, one can check it explicitly that
the resulting scalar potential would not
produce a supersymmetric minimum with a nonzero value of the gluino
condensate. Neither would it yield the correct $Z_{N_c}$ invariance. 
Thus, in order to overcome these difficulties, one  would 
postulate the relation analogous to (\ref {chi}) and 
declare $X$ as the field with respect to  which the variation of
the superpotential should be taken\footnote[2]{Modulo the fact
that both $\chi$ and $X$ are dimensionless fields and while deriving
scalar potentials one should always be working in terms of rescaled
fields with an appropriate  dimensionality.}.  
Thus, one concludes that the relation (\ref {chi}) is a result of 
two symmetries:  $Z_{N_c}$ and supersymmetry. 

Before we turn to the next section it is crucial to present yet 
another form of the superpotential (\ref {super}). One can rewrite it
as 
\begin{eqnarray}
{\cal W}(S, M(X))\equiv \gamma~ S~{\rm ln}{S\over e M(X)} ,
\label{super2}
\end{eqnarray}
where the chiral superfield $M(X)$ is defined in terms of $X$  
\begin{eqnarray}
M(X)\equiv \mu^3 {\rm exp}\Big (-X+{1\over N_c}{\rm sinh}(N_cX) \Big).
\label{M}
\end{eqnarray}
The superpotential (\ref {super2}) has the VY form. 
Since the $X$ field is a dimensionless
field with zero $R$ charge, the expression (\ref {super2}) is
consistent with all the requirements of the VY construction \cite
{VY}. The only difference
is that the scale parameter of the theory $\mu$ is promoted into 
some chiral superfield $M$. The relation between $M$ and $\mu$ 
is such that the $X$ field can be regarded as a dynamical field
setting the value of the phase of the gluino condensate.
This  is reminiscent to the case when the VEV of some dynamical field of 
a bigger theory can be regarded as a parameter in some 
effective theory approximation. 
\vspace{0.4in} \\
{\bf  2.  The General Solution}
\vspace{0.2in}

In this section we derive the general expression for the 
superpotential which satisfies  to the properties listed in the 
previous section.
Our ultimate goal would be to set an expression 
in the form (\ref {super2}), where the dependence of $M$ on $X$
would  be given by a general function compatible with 
the conditions of the problem. 

Thus, we are looking for a 
superpotential in the following form
\begin{eqnarray}
{\cal W}(S,X) =\gamma~S~{\rm ln}{S\over e\mu^3}~+~\gamma~S~{\cal F}(X).
\nonumber
\end{eqnarray} 
This expression can be written as $S$ times some natural logarithm ( as 
in (\ref {super2})). The $S$ superfield has $R$ charge equal to 2
and the mass dimension equal to 3. 
Thus, the $X$ field 
is a dimensionless field with zero $R$ charge. 

One requires that the
superpotential is invariant under the discrete $Z_{N_c}$ 
transformations. Under these
transformations the VY part of the superpotential produces the term
discussed in the previous section. The function ${\cal F}(X)$  should be
chosen in such a way that it would allow  one to eliminate that
term;  i.e.
there should exist a shift of the variable with the following property
\begin{eqnarray}
{\cal F}(X+{\rm shift}) = {\cal F}(X)+i {2\pi k\over N_c}, ~~~~~~~
k=0,~1,...N_c-1. 
\nonumber
\end{eqnarray} 
On the other hand,  it should be possible to perform  
discrete shifts only; indeed, under any continuous $U(1)_R$
transformation the superpotential will be producing 
the anomaly expression, this expression can not be 
eliminated.
In other words, it should be allowed to ``undo'' 
the discrete transformations of $S$ by shifting  $X$,
however, the continuous transformations of $S$  
should not be possible to be eliminated. 
As we discussed it in the previous section 
not all of the shifts of variables are allowed. 
Shifted fields should satisfy appropriate boundary conditions. Thus, 
as before, the shifts are to be  transforming values of the $X$ field 
from one vacuum state to another. 
We will make sure that this is the case here.  
Thus,  the shifts we are looking for 
are of the form 
\begin{eqnarray}
{\cal F}(X+ a \cdot k) = {\cal F}(X)+i {2\pi k\over N_c}, ~~~~~~~
k=0,~1,...N_c-1, 
\label{A1}
\end{eqnarray} 
where $a$ is some nonzero complex number.

The next constraint we are going to impose on the function ${\cal
F}(X)$ is the following. One requires that there are $N_c$ different
minima of the potential. Thus, the equation
\begin{eqnarray}
{\partial {\cal F}(X)\over \partial X}\Big |_{X_k} = 0, 
\label{A2}
\end{eqnarray} 
should have $N_c$ different solutions for $X_k$.  
The shifts in (\ref {A1}) are supposed to  transform  these 
solutions into one another. 

Once the $X$ 
field is integrated out, the resulting 
additional term in the superpotential have to coincide with the term 
(\ref {KSsup}) introduce in Ref. \cite {ShifmanKovner}.
Thus, we get one more condition on the function ${\cal F}$:
\begin{eqnarray}
{\cal F}(X_k) = i~{2\pi k\over N_c}, ~~~~~~~~~~~~~k=0,~1,...N_c-1. 
\label{A3}
\end{eqnarray} 

The solution of Eq. (\ref {A1}) is a sum of its particular
solution and a general solution of the corresponding homogeneous
equation\footnote[8]{This statement is valid for infinitely
differentiable analytic functions.} (let us denote it as ${\cal G}(X)$):
\begin{eqnarray}
{\cal F}(X) =i{2\pi \over N_c a}X ~+~{\cal G}(X).
\label{F1}
\end{eqnarray} 
In terms of the function ${\cal G}(X)$ the expressions (\ref {A1} -- 
\ref {A3})  can be rewritten as
\begin{eqnarray}
{\cal G}(X+a \cdot k) ={\cal G}(X),
\label{A11}
\end{eqnarray} 
\begin{eqnarray}
{\partial {\cal G}(X)\over \partial X}\Big |_{X_k} = -i{ 2\pi\over N_c a}, 
\label{A22}
\end{eqnarray} 
\begin{eqnarray}
{\cal G}(X_k)=- i{ 2\pi\over N_c a}X_k+i{ 2\pi\over N_c}k. 
\label{A33}
\end{eqnarray} 
The solution to (\ref {A11}) is a superposition of
exponential functions which we choose to normalize as follows
\begin{eqnarray}
{\cal G}(X) =-{1\over N_c}\sum_{n=-\infty}^{+\infty}c_n {\rm exp}
\Big (i n {2\pi\over a}X \Big ). 
\label{Gsolution}
\end{eqnarray} 
Substituting Eq. (\ref {Gsolution}) into Eqs. (\ref {A22}) and (\ref {A33})
one derives respectively
\begin{eqnarray} 
\sum_{n=-\infty}^{+\infty}n~c_n {\rm exp}
\Big (i n {2\pi\over a}X_k \Big )=1, 
\label{A222}
\end{eqnarray} 
\begin{eqnarray} 
\sum_{n=-\infty}^{+\infty}c_n {\rm exp}
\Big (i n {2\pi\over a}X_k \Big )=i {2\pi\over a} X_k-i 2\pi k.  
\label{A333}
\end{eqnarray} 
Analyzing Eqs. (\ref {F1}), (\ref {Gsolution}-- \ref {A333})
it is convenient to introduce the following rescaling of the 
$X$ field 
$$
i{2\pi\over N_c a}X \rightarrow X.
$$
Under the shifts discussed above the new variable $X$ transforms as 
\begin{eqnarray} 
X\rightarrow X+i{2\pi\over N_c}k, ~~~~~~~k=0,1,....N_c-1.
\label{Xtransformation}
\end{eqnarray} 
In terms of this variable the expression for the  
function ${\cal F}(X) $ looks as
\begin{eqnarray}
{\cal F}(X)=X~+~{\cal G}(X), 
\label{F2}
\end{eqnarray}
and the expressions  (\ref {Gsolution} -- \ref {A333}) take the form
\begin{eqnarray}
{\cal G}(X) =-{1\over N_c}\sum_{n=-\infty}^{+\infty}c_n {\rm exp}
\Big (n N_c X \Big ), 
\label{Gsolution1}
\end{eqnarray} 
\begin{eqnarray} 
\sum_{n=-\infty}^{+\infty}n~c_n {\rm exp}
\Big (n N_c X_k \Big )=1, 
\label{A2222}
\end{eqnarray} 
\begin{eqnarray} 
\sum_{n=-\infty}^{+\infty}c_n {\rm exp}
\Big (n N_c X_k \Big )=N_c X_k  - i 2\pi k.  
\label{A3333}
\end{eqnarray} 
Let us now consider Eq. (\ref {A2222}). This equation should have
$N_c$ different solutions for $X_k$ describing the $N_c$ vacua. 
It is convenient to introduce 
the notation: $v\equiv {\rm exp}\Big ( N_c X_k \Big )$. In terms of 
$v$ Eq. (\ref {A2222}) could generically 
have an arbitrary big number of solutions.
On the other hand, for each nonzero solution in terms of $v$ there are 
$N_c$ different solutions in terms of $X$; indeed, if the expression
$v=|v| {\rm exp} (i {\rm arg}v)$ is a solution for $v$, then
using the relation  $v= {\rm exp}(N_c X_k )$, one finds $N_c$
different solutions for the imaginary part of $X_k$. 
Thus, in order to have only $N_c$ 
different solutions of Eq. (\ref {A2222}) in terms of $X$, 
the algebraic Eq. 
(\ref {A2222}) should  have  a single nonzero,  multiply 
degenerate root for $v$. 
Let us denote this root as $v\equiv \alpha {\rm exp} (i \rho)$,
with $\alpha$ and $\rho$ being some constants. Then, for the solutions
$X_k$ one gets:
\begin{eqnarray}
{\rm Im } X_k = {\rho \over N_c}~+~{2\pi\over N_c}k, ~~~~~~~
{\rm exp }(N_c  {\rm Re} X_k) = \alpha. 
\nonumber
\end{eqnarray}
One can check now that the shifts  (\ref {Shifts}) really
transform values of $X$ from one vacuum state to some another one. 
For simplicity of arguments, in what follows,  
it is convenient to choose $\alpha =1$ and $\rho =0$.
This corresponds to some shifts of  the $X$
complex coordinate system. 
In that case one derives the following relations
\begin{eqnarray}
 X_k = i {2\pi\over N_c}k,~~~~~\sum_n n~c_n=1,~~~~~~\sum_n c_n=0.  
\label{Xk}
\end{eqnarray}
The particular solution of the previous section corresponds to 
the case when $c_1=1/2$, $c_{-1}=-1/2$ and all other $c$'s are set
to be equal to zero. 

Summarizing, we can write down a general form of the 
superpotential as 
\begin{eqnarray}
{\cal W}(S, M(X))\equiv \gamma~ S~{\rm ln}{S\over e M(X)} ,
\label{super3}
\end{eqnarray}
where the field $M$ is given by the relation 
\begin{eqnarray}
M(X)\equiv \mu^3 {\rm exp}\Big (-{\cal F}(X)\Big ).
\label{M1}
\end{eqnarray}
The function $ {\cal F}(X)$ is defined 
in accordance with Eqs. (\ref {F2}), (\ref {Gsolution1}-- \ref
{A3333}) and the solutions for the vacua in terms of the rescaled 
variable are given in Eq. (\ref {Xk}). 
\vspace{0.2in} \\
{\bf 3.  A Brief Comment  on the K\"ahler Potential}
\vspace{0.2in}

So far we did not discuss what kind of  K\"ahler potential 
$ {\cal K}(S^+S, X^+, X) $ is
supposed to be used in the effective action for $N=1$ SUSY YM model
\footnote[5]{Terms with derivatives of the superfields in 
the K\"ahler potential might lead to an unbounded from below potential
in this case \cite {Shore}. 
For that reason we consider ${\cal K}$ as a function of
the superfields only with no derivatives.}.
There are no symmetry  or anomaly arguments which would 
uniquely fix the form of ${\cal K} $. However, there  is
some piece of information one could  still learn about the  
K\"ahler potential. We would like  to elaborate on this point here. 

In order to make the  K\"ahler potential invariant under 
the shifts  (\ref {Shifts}) one requires that   
${\cal K} (S^+S, X^+, X)$ is actually 
a function of the sum of $X^+$ and
$X$, ${\cal K} (S^+S, X^+, X)={\cal K} (S^+S, X^+ + X)$.
However, this is not the only  form of the expression
which is invariant w.r.t. the shift  (\ref {Shifts}). 
For instance,  the  K\"ahler potential 
could also be  a function of
the sum of ${\cal F}=X+{\cal G}(X)$ and its conjugate ${\cal 
{F}}^+(X^+)$. The sum $X+X^+$ does not change 
upon the shifts (\ref {Shifts})  and the function ${\cal G}$
itself is invariant under those transformations. 
Thus, one could write as a possibility 
$$
{\cal K} (S^+S,~ X^+, X)={\cal K} (S^+S,~ {\cal F}(X)~+~{\cal 
{F}}^+(X^+)). 
$$
We would like to argue here that this type  of dependence of the 
K\"ahler potential is in fact what is dictated by the physical 
particle content of the low-energy spectrum of the 
theory \cite {FGS1}, \cite {FGS2}. 

First let us show that the same combination ${\cal F}(X)~+~{\cal{F}}^+(X^+)$
appears in the expression for the superpotential we derived in the 
previous sections. The part of the superpotential containing 
the chiral superfield ${\cal F}(X)$ is written as $\gamma S {\cal F}(X)$.
This can be presented in the following manner 
\begin{eqnarray}
{\gamma \over N_c} \sum_{k=0}^{N_c-1}~{\cal F}(X)~ \Big (~S-\langle
S\rangle _k~\Big ). 
\label{Formula}
\end{eqnarray}
Then one introduces a real superfield $U_k$ \cite {Gates}
\begin{eqnarray}
U_k=B+i\theta \chi -i {\bar \theta} {\bar \chi}+{ \theta^2  \over 16}
({A^*}
- \langle S\rangle_k^*)+
{ {\bar \theta}^2  \over 16} (A-\langle S\rangle_k)+
{ \theta \sigma^\mu {\bar\theta}\over 48 }
\varepsilon_{\mu\nu\alpha\beta}C^{\nu\alpha\beta}+ 
\nonumber \\
{1\over 2} \theta^2 {\bar \theta} \left ( {\sqrt{2} \over 8}{\bar
\Psi} +{\bar \sigma}^\mu \partial_\mu \chi \right )+
{1\over 2}{\bar  \theta}^2 \theta  \left ( {\sqrt{2} \over 8}
\Psi - \sigma ^\mu \partial_\mu {\bar \chi }\right )+{1\over 4}
\theta^2 {\bar \theta^2} \left ( {1\over 4} \Sigma -\partial^2 B\right), 
\label{UU}
\end{eqnarray} 
which is related to the superfield $S$ 
\begin{eqnarray}
S- \langle S\rangle_k = -4 {\bar D}^2 U_k. 
\label{U}
\end{eqnarray}
The $F$ term of the chiral supermultiplet $S$
is related to the fields $\Sigma$ and $C_{\mu\nu\alpha}$
in the following way\footnote[3]{In this  notation
$\Sigma$ is proportional to $G_{\mu\nu}^2$ and 
$\varepsilon_{\mu\nu\alpha\beta}\partial^{\mu}
C^{\nu\alpha\beta}$ is proportional to $G_{\mu\nu}{\tilde
G^{\mu\nu}}$ \cite{FGS1}.}
\begin{eqnarray}
F=\Sigma+i{1\over 6}\varepsilon_{\mu\nu\alpha\beta}\partial^{\mu}
C^{\nu\alpha\beta}, \nonumber
\end{eqnarray} 
and $A$ and $\Psi$ are respectively the scalar and fermion
components of the superfield $S$. 

One substitutes the expression for $S$ in terms of $U_k$ 
into Eq.  (\ref {Formula}). Then one  
replaces the  ${\bar D}^2$ operator by the integration w.r.t. 
the  $\theta$ variable.  Finally, 
putting the resulting expression together
with its hermitian conjugate part  one derives
\begin{eqnarray}
{\gamma \over N_c} \sum_{k=0}^{N_c-1}~{\cal F}(X)~ \Big (~S-\langle
S\rangle _k~\Big )\Big |_F~+~{\rm h.c.}= {16\gamma \over N_c }~({\cal
F}(X)~+~{\cal {F}}^+(X^+)) ~\sum_{k=0}^{N_c-1} U_k \Big |_D.
\nonumber
\end{eqnarray}

Thus, all the  terms in the Lagrangian of the model 
containing the chiral superfield ${\cal F}(X)$  depend actually on the 
real combination ${\cal F}(X)~+~{\cal {F}}^+(X^+)$. 
This combination can be integrated out using  equations of motion
\cite {Gates}, \cite {FGS2}.
The equation of motion for the real superfield ${\cal F}(X)~+~{\cal 
{F}}^+(X^+)$
leads to the following relation
\begin{eqnarray}
{16\gamma \over N_c}~\sum_{k=0}^{N_c-1}U_k = 
- {\partial {\cal K} (S^+S, Z)\over \partial Z}
\Big |_{Z={\cal F}(X)~+~{\cal {F}}^+(X^+)}.
\label{EOM}
\end{eqnarray}
Thus, the whole Lagrangian can in principle 
be presented in terms of the degrees of
freedom of the real tensor supermultiplet  $U_k$. Indeed, Eq. (\ref {U})
sets how the components of $S$ are related to some components of 
$U_k$, and likewise, Eq. (\ref {EOM}) gives the relation between the 
components of the chiral superfield ${\cal F}$ (or $X$) 
and the components of the superfield $U_k$. This is in agreement with the
statement of Ref. \cite {FGS1} where it was shown that all the 
lowest-spin physical degrees of freedom of SUSY gluodynamics
can be described by one  real tensor supermultiplet $U\equiv 
(\sum_{k=0}^{N_c-1}U_k)/N_c$. One should notice that once 
the ${\cal F}$ field, being appropriately rescaled, is considered
as an  independent fundamental field of the Lagrangian for which Eq. 
(\ref {EOM}) is to be solved, the whole information on the $Z_{N_c}$
vacuum structure is lost and one is simply dealing with
some particular ground state. It is the definition of ${\cal F}$
in terms of $X$ that makes the $Z_{N_c}$ structure feasible  and Eq. 
(\ref {EOM}) should actually be solved for the $X$ 
field being appropriately rescaled. In other words, the information
about the $Z_{N_c}$ structure in this case is encoded in the relations
between $U_k$ and $S$ and $U_k$ and $X$. 

We conclude that 
the form of the K\"ahler potential which is dictated by the 
particle content of the model in some particular vacuum state \cite
{FGS1}, \cite {FGS2} is consistent with the symmetry 
requirements we have used to 
derive the effective superpotential in the previous sections.  
\vspace{0.2in} \\
{\bf Discussions}
\vspace{0.2in}

We derived the effective superpotential for SUSY gluodynamics
which correctly reproduces the known properties 
of the complicated ground state structure of the model. The corresponding
scalar potential is a smooth function of arguments and 
yields  $N_c$ different vacua with the broken chiral invariance. The discrete
$Z_{N_c}$ transformations shift one vacuum state into another one.
The superpotential is given  in terms of two chiral
superfields. Once one superfield is integrated out, the superpotential
reduces to the expression given in  Ref. \cite {ShifmanKovner}. 
On the other hand, 
if one is restricted to study the excitations about 
some particular vacuum state
with the nonzero gluino condensate only, then 
the superpotential reduces to the 
known expression of Ref. \cite {FGS2}. This last adequately describes 
all the lowest-spin degrees of freedom of the model \cite {FGS2}.
The superpotential (\ref {super}) can
formally be brought to the original VY logarithmic 
form (see Eqs. (\ref {super2}) and (\ref {super3})).
In this case one could  think of the VY superpotential where
the scale parameter of the model $\mu$ is promoted into 
some dynamical chiral 
superfield. The VEV of the phase of that superfield would
set the value of the phase of the gluino condensate. This 
superfield, as we have shown, is related to pure gluonic operators. 
Finally, as we mentioned before,  the  superpotential can be used to
study the domain walls separating the chirally asymmetric vacua of the
theory. The results of those studies will be reported elsewhere. 
\vspace{0.2in} \\
{\bf Acknowledgments}
\vspace{0.1in} 

The author is grateful to G.R. Farrar, Z. Kakushadze, M. Schwetz
for reading the manuscript and for useful discussions and 
suggestions. Discussions with A.V. Smilga are also very appreciated.

\end{document}